\documentstyle[]{mn}

\newcommand{\etal}{\mbox{et\ al.\ }}
\newcommand{\kmsec}{\,\mbox{$\mbox{km}\,\mbox{s}^{-1}$}}

\newcommand{\hb}{\hbox{$\hbox{H}\beta$}}
\newcommand{\hg}{\hbox{$\hbox{H}\gamma$}}

\newcommand{\heii}{\hbox{$\hbox{He\,{\sc ii}\,$\lambda$4686\,\AA}$}}

\title[Atlas of dwarf novae in outburst]
      {Spectral atlas of dwarf novae in outburst}
\author[L.\ Morales-Rueda, T.\, R.\ Marsh]
       {L.\ Morales-Rueda, T.\, R.\ Marsh \\
       Department of Physics and Astronomy, Southampton University,
      Southampton SO17 1BJ, UK\\ 
       (lmr@astro.soton.ac.uk, trm@astro.soton.ac.uk)\\}
\date{Accepted ...
      Received ...;
      in original form ...}

\begin{document}

\maketitle

\begin{abstract}
  
  Up to now, only a very small number of dwarf novae have been studied
  during their outburst state ($\sim$30 per cent in the Northern
  hemisphere). In this paper we present the first comprehensive atlas
  of outburst spectra of dwarf novae. We study possible correlations
  between the emission and absorption lines seen in the spectra and
  some fundamental parameters of the binaries. We find that out of the
  48 spectra presented, 12 systems apart from IP~Peg show strong
  He{\sc ii} in emission: SS~Aur, HL~CMa, TU~Crt, EM~Cyg, SS~Cyg,
  EX~Dra, U~Gem, HX~Peg, GK~Per, KT~Per, V893~Sco, IY~UMa, and 7
  others less prominently: FO~And, V542~Cyg, BI~Ori, TY~Psc, VZ~Pyx,
  ER~UMa, and SS~UMi. We conclude that these systems are good targets
  for finding spiral structure in their accretion discs during
  outburst if models of Smak (2001) and Ogilvie (2001) are correct.
  This is confirmed by the fact that hints of spiral asymmetries have
  already been found in the discs of SS~Cyg, EX~Dra and U~Gem.

\end{abstract}
 

\section{Introduction}

Dwarf novae are a type of cataclysmic variable that undergo outbursts
during which they increase in brightness by 2--8 mags. The outburst
recurrence time varies widely from system to system and can be as
short as a few days (ER~UMa systems) and as long as tens of years
(WZ~Sge systems). The most likely mechanism for dwarf nova outbursts
is a thermal instability within an accretion disc (Osaki 1974). There
are still many unknowns in the disc instability theory. For example,
although we know that the viscosity in the disc plays a fundamental
role in driving the outburst, we do not know what its origin is. In
1986 Sawada, Matsuda \& Hachisu suggested that the disc viscosity
could be due to shocks in the disc tidally excited by the donor star.
In 1991, Balbus \& Hawley discovered a strong magneto-hydrodynamical
instability of accretion discs which can indeed produce considerable
effective viscosity. But it is not yet certain which, if either, of
these is dominant.

There are several extensive studies on dwarf novae in the literature
(Warner 1995 and references therein) but they concentrate mainly on
their quiescent states.  Only $\sim$30 per cent of known Northern
hemisphere dwarf novae have published outburst spectra and the number
for Southern hemisphere systems is even smaller. The main aim of this
paper is to fill this gap by presenting spectra of 48 dwarf novae
during their outburst state. We carry out comparisons between the
spectra of the different dwarf novae and search for possible
correlations between the features seen in the spectra and fundamental
parameters of the systems such as their inclination, the mass of their
components, their orbital periods and the outburst phases at which
they were observed. Comparisons between the spectra are possible
because the data were taken with comparable setups during a long term
programme using service time.

Another reason for compiling this atlas is to look for spectra similar
to that of IP~Peg, U~Gem and WZ~Sge, systems that have been confirmed
to show spiral structure in their discs during outburst. By confirmed
we mean that spiral structure has been detected in more than one set
of data, either taken during a different outburst or during the same
outburst at a different time (IP~Peg: Steeghs et al. 1997, Harlaftis
et al. 1999, Morales-Rueda, Marsh \&\ Billington 2000; U~Gem: Groot
2001, Steeghs private communication; WZ~Sge: Kuulkers et al. 2001,
Baba, Sadakane \&\ Norimoto 2001). The principal feature of their
spectra, compared with those of other dwarf novae, is that they show
strong lines in emission instead of absorption. Of the $\sim$30 per
cent of dwarf novae with outburst spectra published, only half of
these show any emission, and even then it is often in the form of
emission lines buried in absorption troughs. There could easily be
systems with strong emission lines in outburst that have yet to be
identified. Of most interest is the presence of \heii\ in emission as
this line is very important from the irradiation point of view. If
these asymmetric structures seen in the discs of IP~Peg, U~Gem and
WZ~Sge are the result of tidally thickened sectors of the disc being
irradiated by the white dwarf, boundary layer and/or inner disc during
the outburst (as suggested by Smak 2001 and Ogilvie 2001) we expect
them to be seen most clearly in \heii.

\section{Observations}
 \begin{table*}
 \centering
 \caption{List of the dwarf novae (DNe) in outburst observed during
   this program. A note is added when the dwarf nova is of the SU~UMa or Z~Cam
   type (shows superoutbursts or standstills as well as normal outbursts). The
   date when the system was observed is given as well as its
   orbital period, P$_{\rm orb}$, when known (Ritter \& Kolb 1998, (a)
   Thorstensen 1997, (b) Uemura et al. 2000). All DNe were observed
   using the INT except IR~Gem and IY~UMa which were observed with the
   WHT.}
 \label{obs:jo}
 \begin{center}
 \begin{tabular}{lclllcll}
Target & DNe type & Date & P$_{\rm orb}$ (d) & Target &
DNe type & Date & P$_{\rm orb}$ (d)\\
& & observed & & & & observed &\\
\hline
FO~And    & suuma & 2000/10/17 & 0.0716 &IR~Gem    & suuma & 2000/01/18 & 0.0684 \\
LX~And    &       & 2000/08/15 &        &U~Gem     &       & 2000/03/02 & 0.1769 \\
RX~And    & zcam  & 2000/10/21 & 0.2099 &X~Leo     &       & 1999/10/21 & 0.1644 \\
FO~Aql    &       & 2000/08/15 &        &CY~Lyr    &       & 2000/03/14 & 0.1591 \\
VZ~Aqr    &       & 1999/07/16 &        &V419~Lyr  & zcam  & 1999/08/26 &        \\
SS~Aur    &       & 2001/03/07 & 0.1828 &BI~Ori    &       & 2000/03/02 &        \\
AT~Cnc    & zcam  & 2000/03/14 & 0.2387 &CN~Ori    & zcam  & 2000/03/14 & 0.1632 \\
SY~Cnc    & zcam  & 2001/03/07 & 0.3800 &V1159~Ori & suuma & 2000/03/14 & 0.0622\\
YZ~Cnc    & suuma & 2000/03/14 & 0.0868 &HX~Peg    &       & 2000/07/03 & 0.2008\\
HL~CMa    & zcam  & 2000/03/14 & 0.2145 &IP~Peg    &       & 1994/08/30 & 0.1582\\
SV~CMi    & zcam  & 2000/10/17 & 0.1560 &RU~Peg    &       & 1999/08/19 & 0.3746\\
AM~Cas    & zcam  & 2000/10/17 & 0.1650 &GK~Per    &       & 1996/02/28 & 1.9968\\
GX~Cas    & suuma & 1999/10/21 & 0.0890 &KT~Per    & zcam  & 1999/10/31 & 0.1627\\
KU~Cas    &       & 2000/11/6  &        &TZ~Per    & zcam  & 1999/08/19 & 0.2605\\
TU~Crt    & suuma & 2001/04/11 & 0.0844 &TY~Psc    & suuma & 2000/11/6  & 0.0683\\
EM~Cyg    & zcam  & 1999/08/26 & 0.2909 &VZ~Pyx    & suuma & 2001/03/07 & 0.0740\\
SS~Cyg    &    & 1999/08/19,26 & 0.2751 &AW~Sge    &       & 2000/07/16 &   \\
V516~Cyg  &       & 2000/10/17 &        &RZ~Sge    & suuma & 2000/10/17 & 0.0686\\
V542~Cyg  &       & 2000/03/14 &        &V893~Sco  &       & 2000/08/15 & 0.0761\\
V792~Cyg  &       & 2000/10/17 &        &NY~Ser    & suuma & 2000/04/10 & 0.1007\\
V1504~Cyg & zcam  & 2000/07/03 & 0.0695 &DI~UMa    &       & 2000/03/02 & 0.0548\\
AB~Dra    & zcam  & 1999/08/26 & 0.1520 &ER~UMa    & suuma & 2000/11/6  & 0.0637\\
EX~Dra    &       & 2001/05/09 & 0.2099 &IY~UMa    & suuma & 2000/01/18 & 0.0738$^{b}$\\ 
AH~Eri    &       & 2000/11/6  & 0.2391$^{a}$ &SS~UMi    & suuma & 2000/08/15 & 0.0678\\
\hline
 \end{tabular}
 \end{center}
 \end{table*}
 
 The data presented in this paper were taken by different support
 astronomers during service time. All the spectra but two were
 obtained with the Intermediate Dispersion Spectrograph (IDS) mounted
 on the 2.5\,m Isaac Newton telescope (INT) on La Palma. The spectra
 of IR~Gem and IY~UMa were taken with the blue arm of ISIS on the 4\,m
 William Herschel telescope (WHT). Two different IDS setups were used
 for the observations. The first setup consisted of the 235\,mm camera
 with the R1200B grating. The second setup used the 500\,mm camera and
 the R632V grating. The observations were taken with one of the two
 CCDs available with the IDS, either the thinned EEV10 (2Kx4K) or the
 Tek (1Kx1K). Both setups produced almost the same wavelength
 dispersion (35 and 31 \AA\ mm$^{-1}$ respectively).  In both cases we
 covered \hg\ and \heii\ and a few He\,{\sc i} lines.  The two spectra
 taken with the WHT used the blue arm of ISIS with the R1200B grating
 and a EEV CCD covering the same wavelength region as the INT spectra
 with a wavelength dispersion of 33 \AA\ mm$^{-1}$.  Exposure times
 were generally 600\,s.  Table~\ref{obs:jo} gives a list of the dwarf
 novae observed.
 
 The images were debiased and flat-fielded using bias and tungsten
 lamp exposures respectively. The optimal extraction algorithm of
 Horne (1986) was used to extract the spectra. The data were
 wavelength calibrated using a CuAr arc lamp. The arcs were extracted
 using the profile determined for their corresponding stellar image to
 avoid possible systematic errors due to tilted spectra. We observed
 one flux standard with each of the setups: BD+28$^{\circ}$4211 (Oke
 1990) with the first setup on 2000 August 15, BD+33$^{\circ}$2642
 (Oke 1990) with the second setup on 2000 April 10, and HD19445 (Oke
 \& Gunn 1983) with the WHT on 2000 January 18.  These were used to
 correct the data from instrumental response and extinction.

\section{Results}

\begin{figure*}
\begin{picture}(100,0)(-270,250)
\put(0,0){\includegraphics{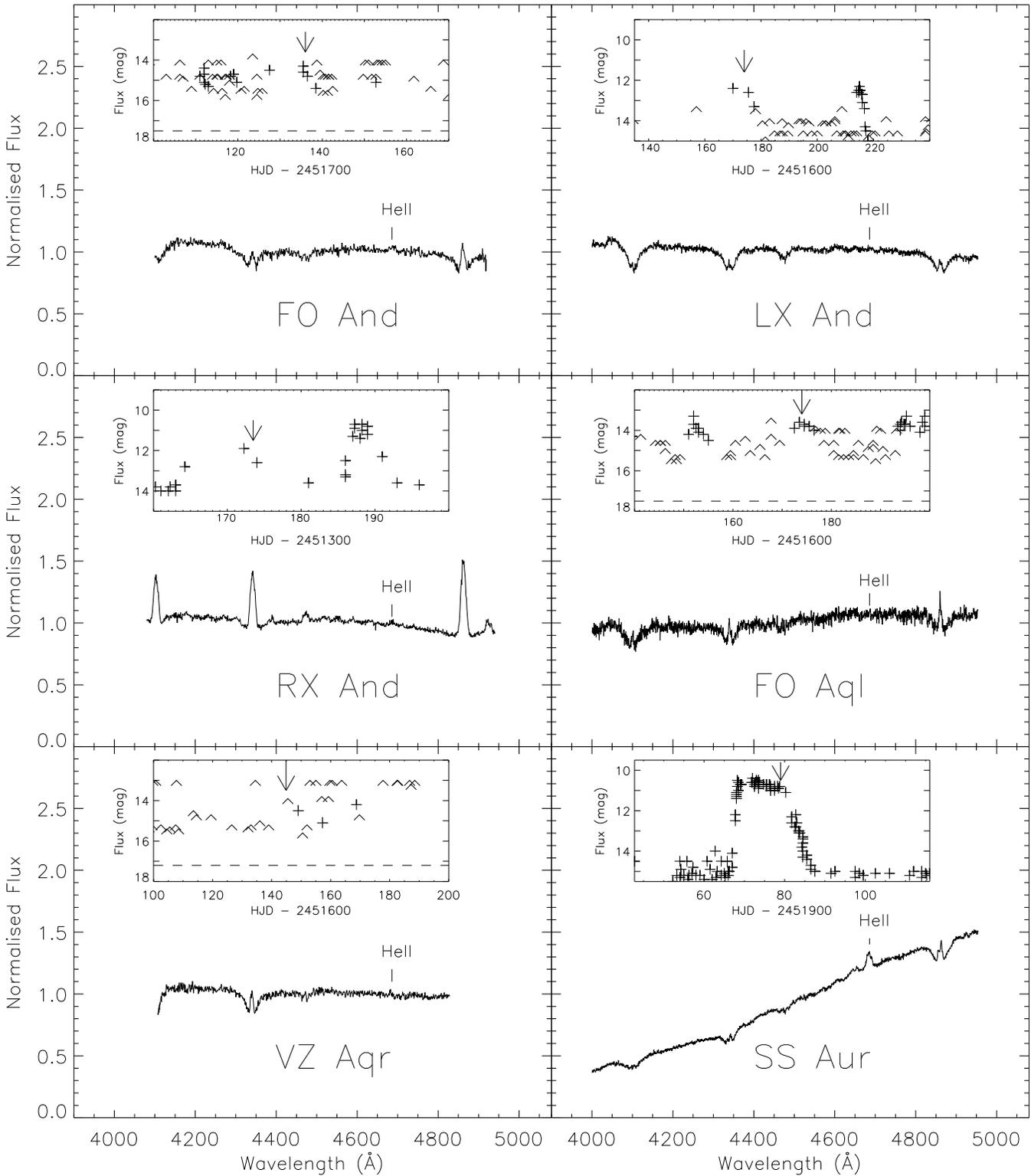}}
\noindent
\end{picture}
\vspace{210mm}
\caption{Outburst spectra of 6 dwarf novae. Each panel contains the spectrum
  of the object and another panel with its lightcurve. An arrow
  indicates the time when the spectrum was taken. A horizontal dashed
  line indicates the magnitude of the quiescent level as given by
  Downes et al. (1997). The normalised flux is plotted from zero to
  show the relative line strengths with respect to the continuum. See
  text for more details.}
\label{res:avsp1}
\end{figure*}

\begin{figure*}
\begin{picture}(100,0)(-270,250)
\put(0,0){\includegraphics{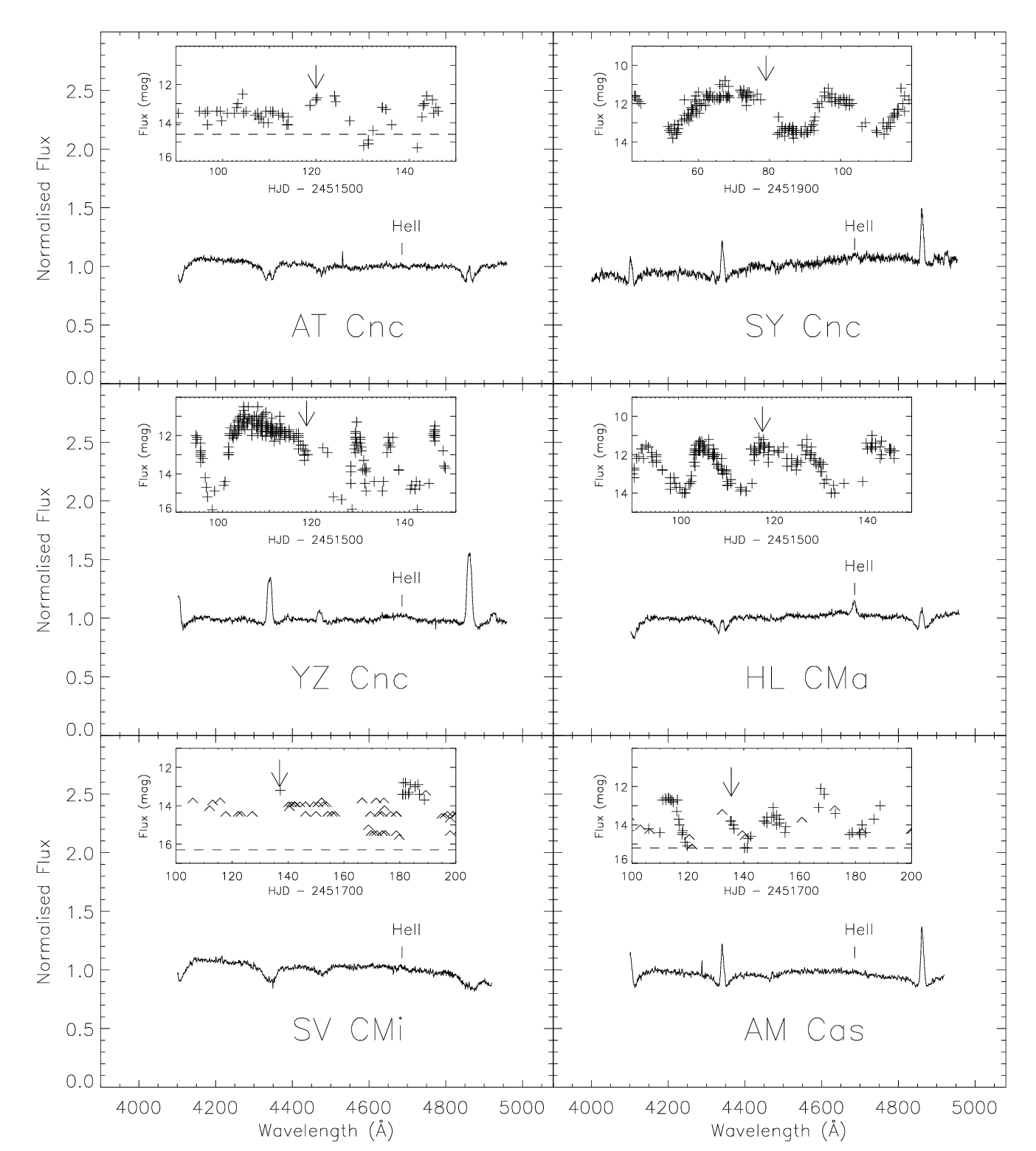}}
\noindent
\end{picture}
\vspace{210mm}
\caption{Same as in Fig.~\ref{res:avsp1}.}
\label{res:avsp2}
\end{figure*}

\begin{figure*}
\begin{picture}(100,0)(-270,250)
\put(0,0){\includegraphics{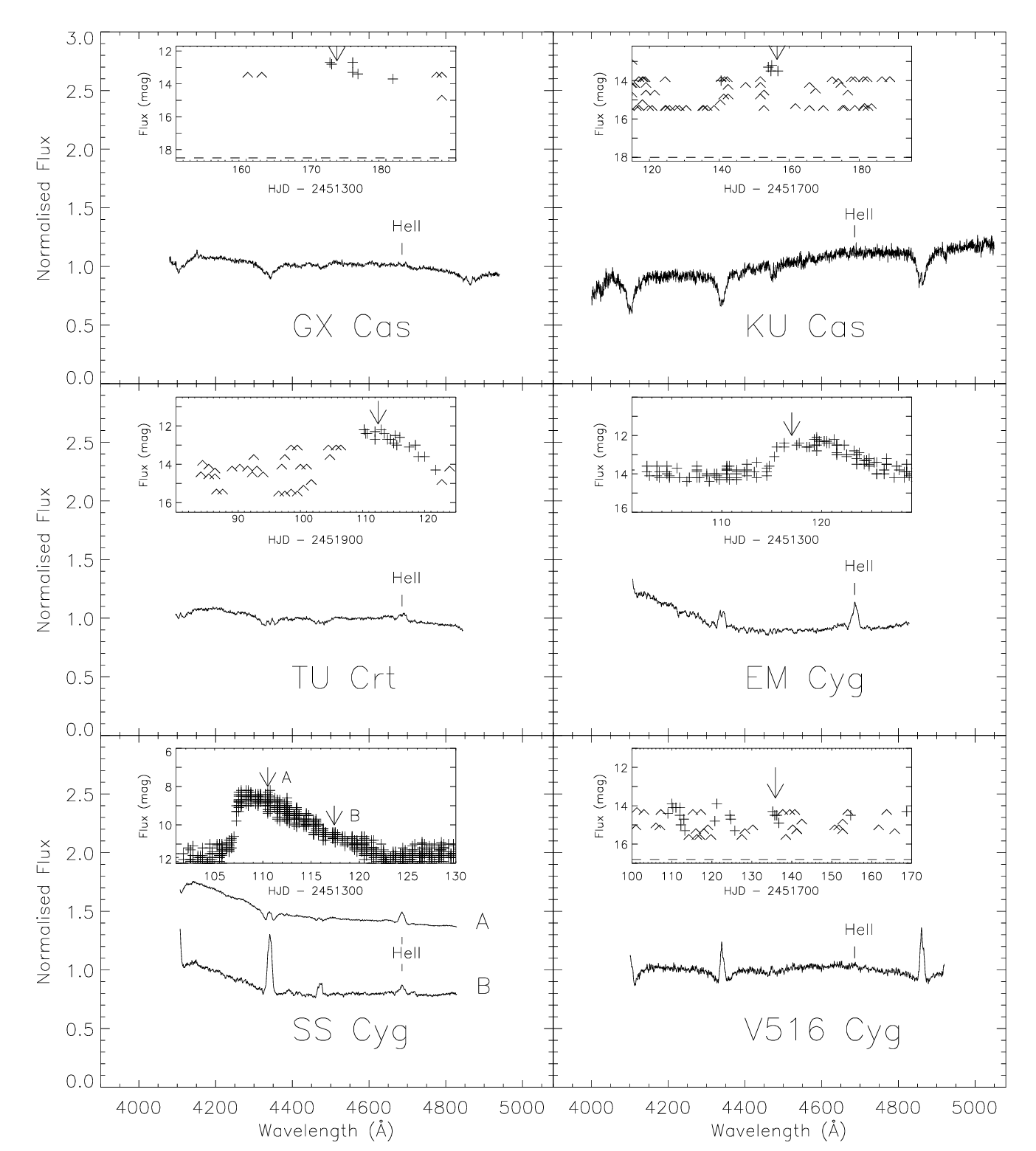}}
\noindent
\end{picture}
\vspace{210mm}
\caption{Same as in Fig.~\ref{res:avsp1}.}
\label{res:avsp3}
\end{figure*}

\begin{figure*}
\begin{picture}(100,0)(-270,250)
\put(0,0){\includegraphics{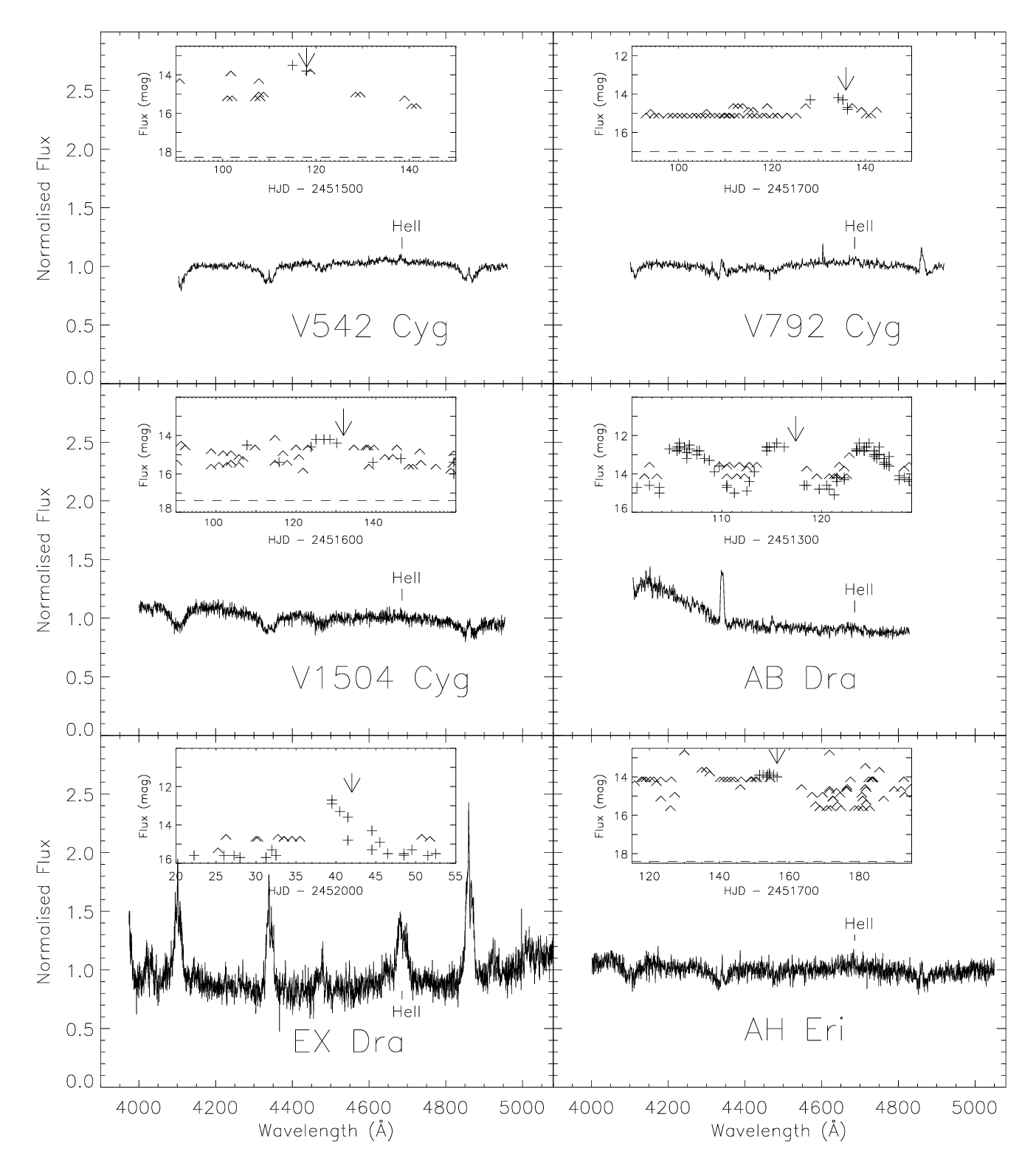}}
\noindent
\end{picture}
\vspace{210mm}
\caption{Same as in Fig.~\ref{res:avsp1}.}
\label{res:avsp4}
\end{figure*}

\begin{figure*}
\begin{picture}(100,0)(-270,250)
\put(0,0){\includegraphics{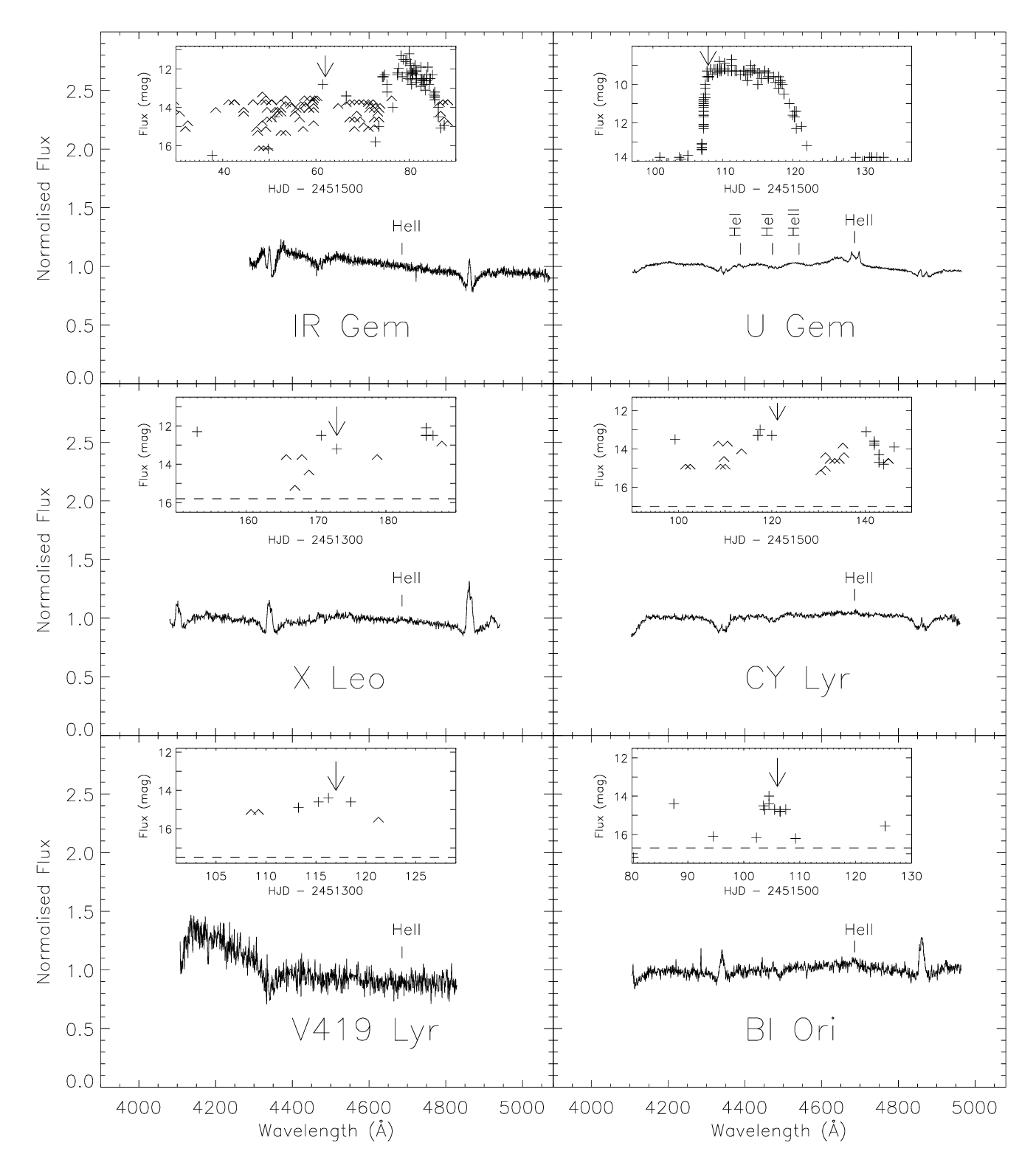}}
\noindent
\end{picture}
\vspace{210mm}
\caption{Same as in Fig.~\ref{res:avsp1}.}
\label{res:avsp5}
\end{figure*}

\begin{figure*}
\begin{picture}(100,0)(-270,250)
\put(0,0){\includegraphics{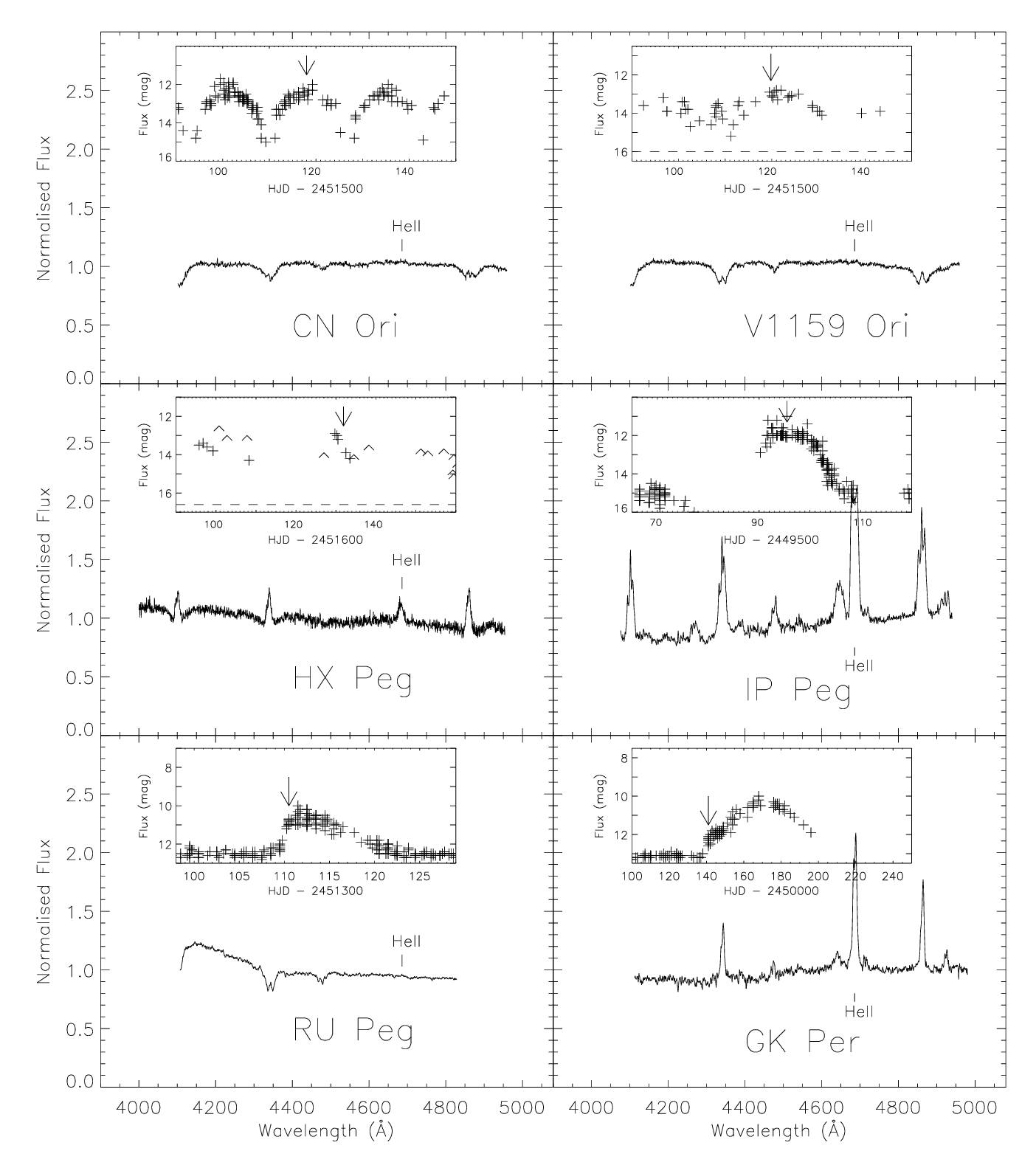}}
\noindent
\end{picture}
\vspace{210mm}
\caption{Same as in Fig.~\ref{res:avsp1}.}
\label{res:avsp6}
\end{figure*}

\begin{figure*}
\begin{picture}(100,0)(-270,250)
\put(0,0){\includegraphics{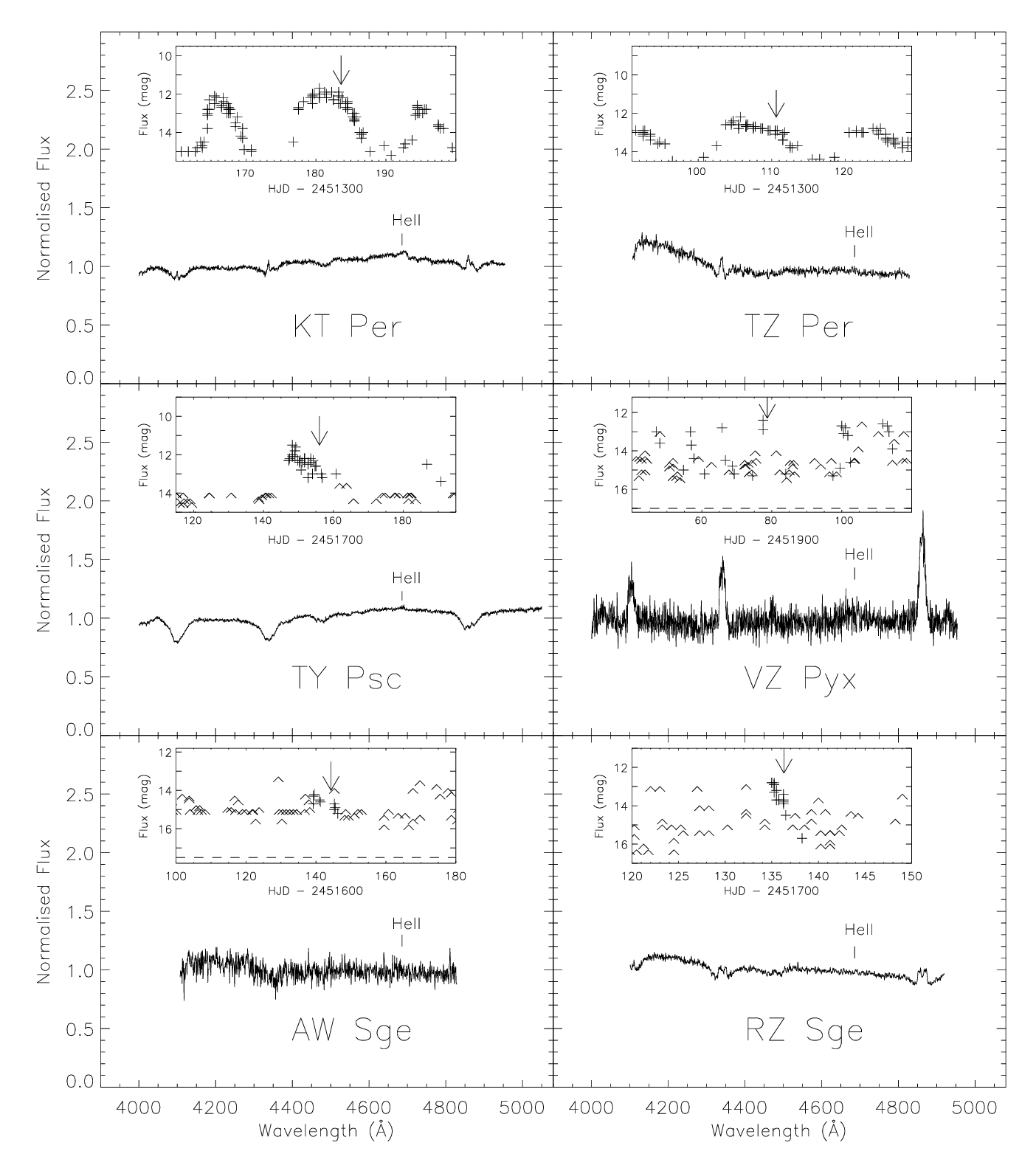}}
\noindent
\end{picture}
\vspace{210mm}
\caption{Same as in Fig.~\ref{res:avsp1}.}
\label{res:avsp7}
\end{figure*}

\begin{figure*}
\begin{picture}(100,0)(-270,250)
\put(0,0){\includegraphics{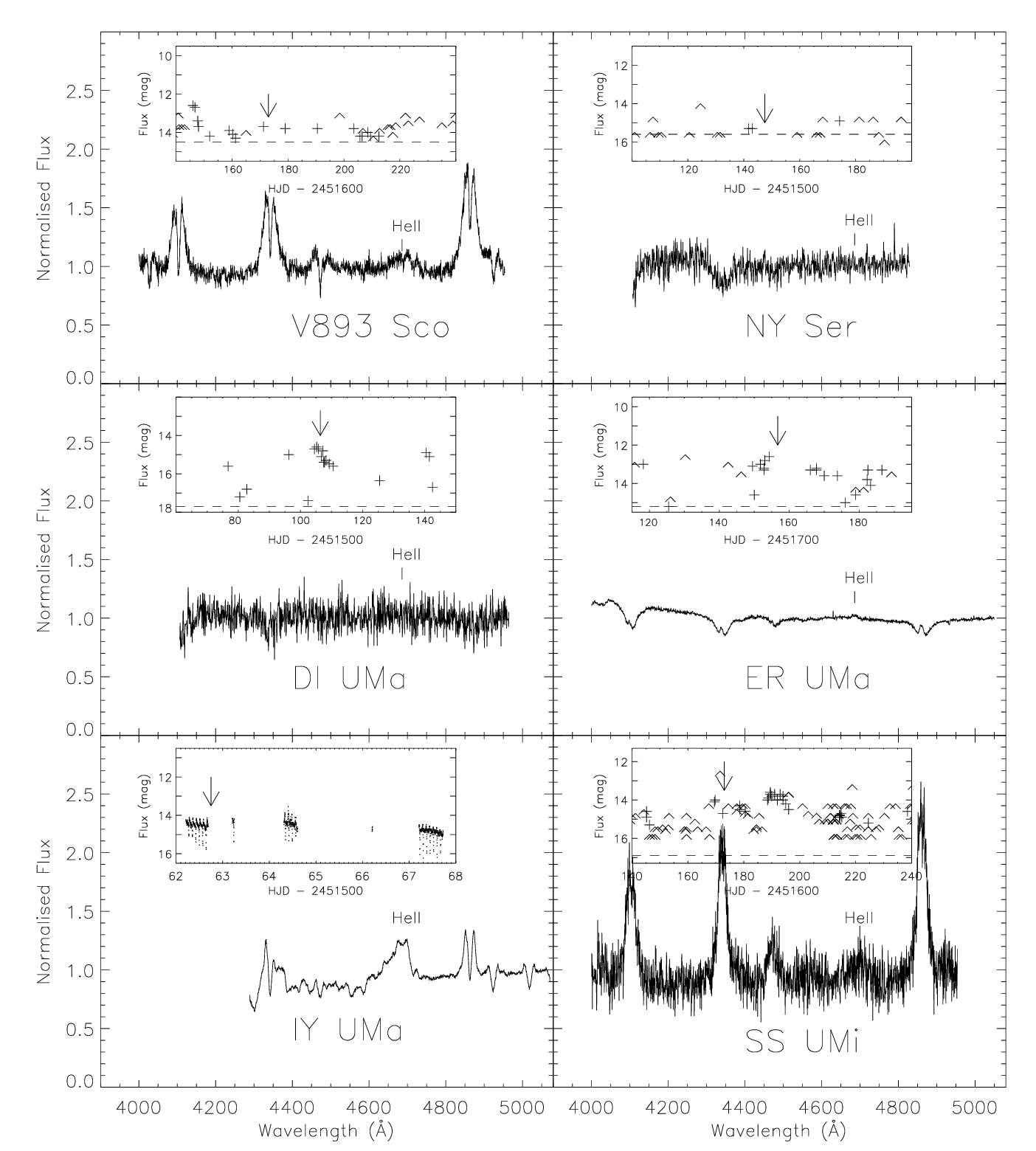}}
\noindent
\end{picture}
\vspace{210mm}
\caption{Same as in Fig.~\ref{res:avsp1}.}
\label{res:avsp8}
\end{figure*}

Figs. \ref{res:avsp1} to \ref{res:avsp8} present the spectra of 48
dwarf novae during their outburst state. Above the spectra there is
another panel showing the VSNET (Variable Star NETwork), VSOLJ
(Variable Star Observers League in Japan) or AAVSO (American
Association of Variable Star Observers) lightcurve with an arrow
pointing at the time when our spectra were taken. For those systems
where there are very few data points in the light curves we have also
plotted a horizontal dashed line indicating the magnitude of the
quiescent state as given by Downes, Webbink \& Shara (1997).  Of all
the dwarf novae presented, only a few were observed on the same night
as the standards used for their flux calibration: FO~Aql, LX~And,
IR~Gem, IP~Peg, GK~Per, V893~Sco, NY~Ser, IY~UMa and SS~UMi. For the
other 39 systems we used the standard taken with the same setup as the
targets to calibrate their fluxes. In these cases the slope of the
continuum should be taken with caution. Most of the spectra show a
blue continuum characteristic of cataclysmic variables. An exception
to this is SS~Aur that seems to show a red excess in the continuum. We
do not believe this to be real but the result of incorrect flux
calibration. We have not been able to track down the problem in this
case. After flux calibration, we divided each spectrum by a constant
obtained by averaging the values of the continuum in regions free of
lines.  We plot all the spectra from zero to show the relative line
strengths.

Table~3 gives the measured equivalent width (EW) and full width at
half maximum (FWHM) of the lines present in the spectra as well as the
inclination and masses of the compact object and the donor star when
known, as taken from Ritter \& Kolb (1998) unless otherwise indicated.
To measure EWs we considered the absorption lines to extend from
$-$3500 \kmsec\ to 3500 \kmsec\ for the Balmer lines and from $-$3000
\kmsec\ to 3000\kmsec\ for the He\,{\sc i} lines. For emission lines
with no sign of absorption we considered the line to comprise from
$-$1000 \kmsec\ to 1000 \kmsec\ of the rest wavelength.  In the case
of \heii\ we tried to avoid contamination by the nearby Bowen blend
and He\,{\sc i}\,$\lambda$4713\,\AA\ by choosing the line limits by
eye in each case. In the particular case of IY~UMa only the EW of \hb\ 
is reliable as \hg\ and \heii\ are heavily contaminated by nearby
lines.  FWHMs were measured by fitting a Gaussian function to the
emission or absorption line. The FWHM, the offset from the rest
wavelength, and the peak of the Gaussian were free parameters of the
fit. In the case of absorption lines with emission cores, the FWHM of
the absorption component were measured by masking the central
emission. In those cases, two values for FWHM are given, one for the
absorption line and another for the central emission. In the case of
emission lines with absorption cores, the absorption was also masked
before the Gaussian fit was attempted but only the emission FWHM is
given.

For comparison we have also included the spectra of GK~Per during its
1996 outburst (Morales-Rueda, Still \&\ Roche 1999) and that of
IP~Peg, during its August 1994 outburst (Morales-Rueda et al. 2000).
The spectrum of IP~Peg is very different from that of the rest of the
dwarf novae, the most striking difference being the strong \heii\ line
in emission.

\subsection{Notes on particular objects}

\quad\quad{\it SS~Aur}. As noted above, the red continuum seen in the
spectra of this system is probably the result of poor flux
calibration.  \vspace{3mm}

{\it SS~Cyg}. We took two spectra of this dwarf nova during the same
outburst, one near maximum, the other in the descent to quiescence.
The spectra at these two times show some major differences. At maximum
the continuum increases significantly. As a result of the disc
becoming optically thick, we see absorption lines with an emission
core. The strength of the \heii\ line is significantly larger relative
to the strength of the Balmer and He\,{\sc i} lines compared to the
quiescent state. In contrast, the spectrum taken near quiescence shows
a lower continuum with Balmer and He emission lines.  SS~Cyg is one of
the systems claimed to have spiral asymmetries during outburst
(Steeghs \etal\ 1996).  \vspace{3mm}

{\it U~Gem}. The spectrum shows Balmer absorption lines with faint
emission cores. \heii\ appears strong in the spectrum and double
peaked presumably due to the high inclination of the system. There is
also some emission in the Bowen blend. In a recent study of U~Gem
during outburst Groot (2001) finds spiral structure in the accretion
disc. Spectra taken on 2001 April 30, and May 4, 5, 9 and 10 when
U~Gem was in outburst (first three nights) and decaying to quiescence
(last two nights) show the presence of spiral structure in \heii\ and
less clearly in \hb\ while the system was in outburst. The \heii\ 
emission line disappears almost completely from the spectrum on May 9
and 10, when the system was half way down to its quiescent state
(Steeghs, private communication). This indicates that the origin of
\heii\ is either irradiation caused by a thickening of the accretion
disc during outburst, as suggested by Ogilvie (2001), or the spiral
asymmetries themselves.

Other features worth mentioning about the spectrum of U~Gem are the
bumpy structures seen to the blue of \heii. These are seen in other
spectra of this sample, i.e. KT~Per, TY~Psc, V516~Cyg. These features
are real and correspond to multiple He\,{\sc i} and Fe\,{\sc ii}
lines. Some of them have been marked in the spectrum of U~Gem for
clarity.  \vspace{3mm}

{\it AW~Sge, NY~Ser, DI~UMa}. These spectra are particularly noisy as
these dwarf novae are still quite faint during outburst and our
standard exposure time did not achieve the intended signal to noise.
The signal-to-noise ratio of these spectra does not allow us to
conclude whether \heii\ is present in emission or not but they clearly
show Balmer lines in absorption, perhaps with central emission. They
probably resemble the spectra of RU~Peg or V1159~Ori more than that of
IP~Peg.  \vspace{3mm}

{\it IY~UMa}. This spectrum shows strong Balmer and He\,{\sc i} lines
in emission with absorption cores (the opposite to what we have seen
up to now). There is strong emission in \heii\ and the Bowen blend.
Although from the light curve we can see that our spectrum does not
coincide with amateur observations, we are certain that IY~UMa was in
outburst when we observed it as its quiescent level is magnitude
$\sim$17. The rapid variations seen in the light curve are due to
eclipses.

\subsection{\heii\ in emission}

Out of the sample of CVs observed we find, apart from IP~Peg, 12 dwarf
novae that show \heii\ in emission: SS~Aur, HL~CMa, TU~Crt, EM~Cyg,
SS~Cyg, EX~Dra, U~Gem, HX~Peg, GK~Per, KT~Per, V893~Sco, IY~UMa, and
possibly 7 other systems were \heii\ is present but very faint:
FO~And, V542~Cyg, BI~Ori, TY~Psc, VZ~Pyx, ER~UMa, and SS~UMi.

\section{Discussion}

\begin{figure}
\begin{picture}(100,0)(-270,250)
\put(0,0){\includegraphics{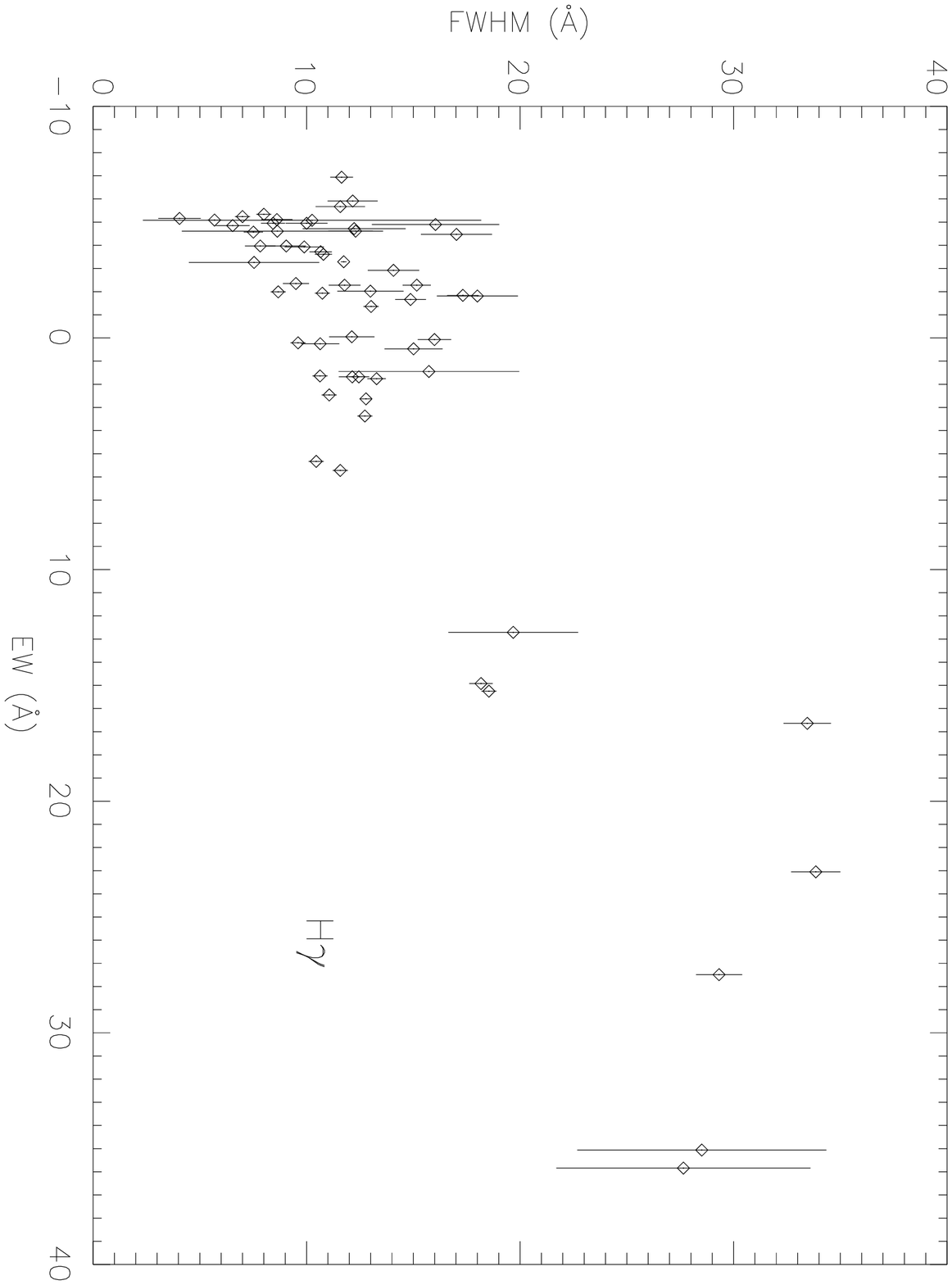}}
\put(0,0){\includegraphics{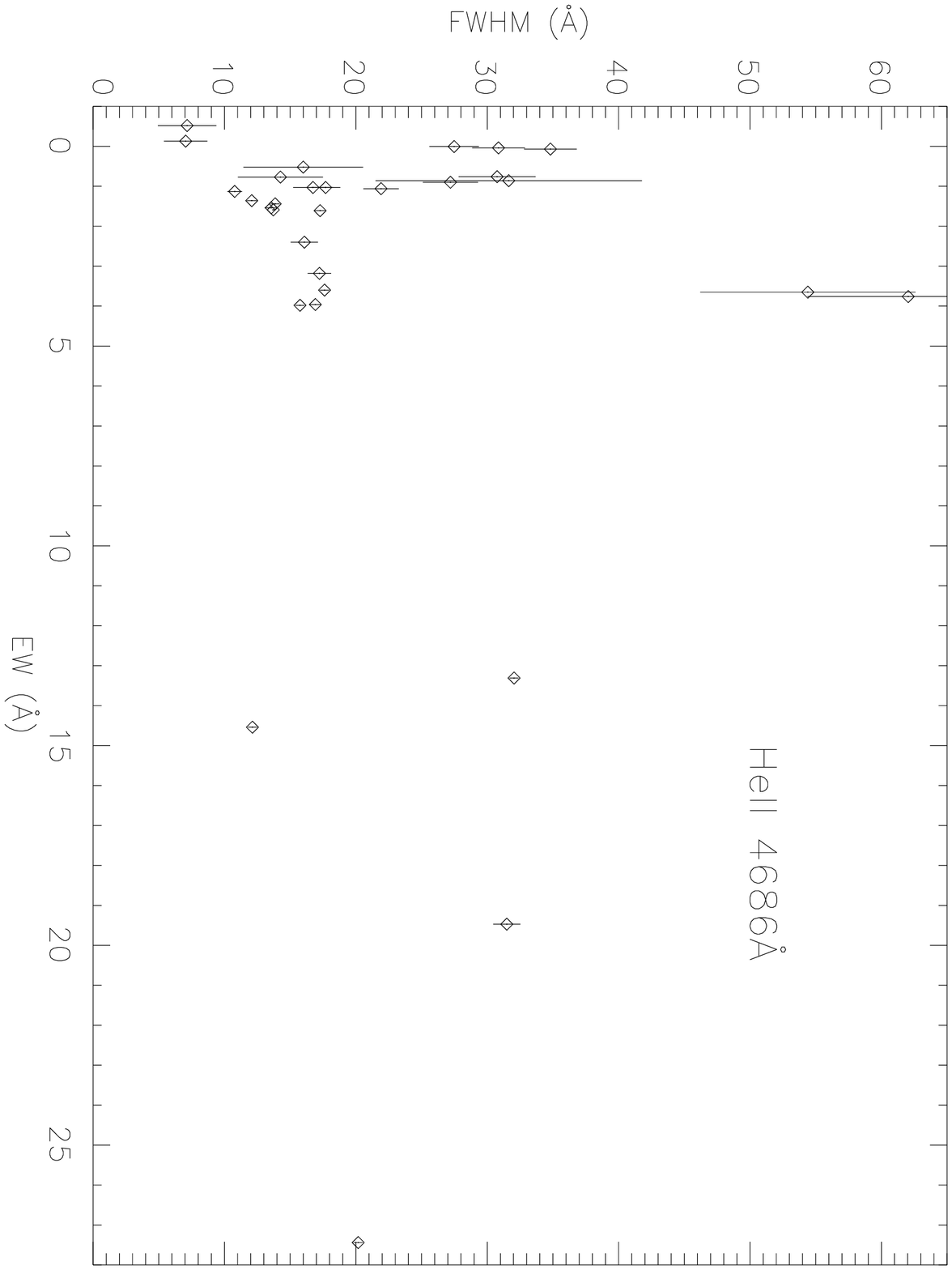}}\noindent
\noindent
\end{picture}
\vspace{140mm}
\caption{FWHM versus EW for \hg\ (top panel) and \heii\ (bottom panel.)}
\label{res:ewfwhm1}
\end{figure}

In this section we look for correlations between the quantities we can
measure in our spectra (i.e. EW, FWHM, outburst phase) and fundamental
parameters of the dwarf novae, like their masses and orbital periods.

\subsection{FWHM versus EW}

For high inclination dwarf novae, the FWHM of the lines scales as
$\sin i$ (where $i$ is the inclination of the system). This is due to
Doppler broadening of the lines. It is not clear that this is true for
low inclination systems. At low inclinations other mechanisms may
dominate over Doppler broadening, e.g. thermal broadening, intrinsic
broadening. It is not possible to establish at what range of
inclinations the FWHM of the lines is directly proportional to $\sin
i$ because we are not capable of measuring $\sin i$ with any certainty
for systems that are not eclipsing.

Keeping this in mind we search for correlations between the FWHM of
the lines and their strength, as given by their EW.
Fig.~\ref{res:ewfwhm1} shows the FWHM versus EW for \hg\ (top
panel) and \heii\ (bottom panel).

There is some indication of positive correlation of FWHM with EW. This
apparent correlation would indicate that high inclination systems
(those with higher FWHM) show stronger emission lines. It is worth
mentioning that three of the systems where \heii\ is strong in
emission during outburst and where spiral structure has been detected,
i.e. IP~Peg, EX~Dra and U~Gem, are edge-on systems and one expects
radiative transfer effects to favour seeing emission in such cases.

\subsection{Line emission versus the masses of the components of the system}

\begin{figure}
\begin{picture}(100,0)(-270,250)
\put(0,0){\includegraphics{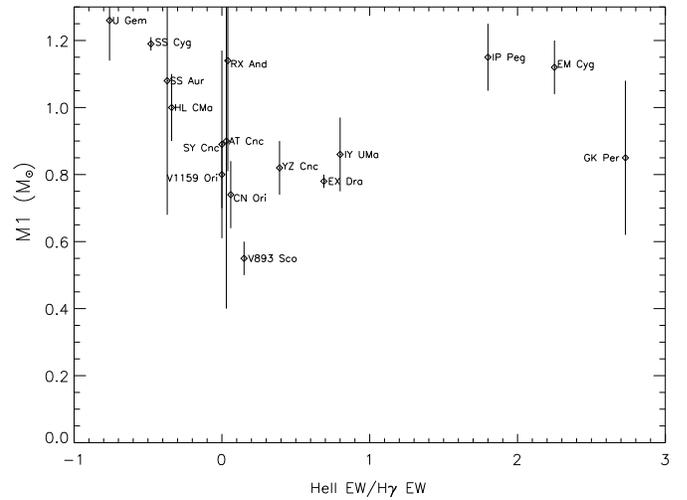}}
\noindent
\end{picture}
\vspace{70mm}
\caption{Mass of the white dwarf in the dwarf nova versus the ratio of
  equivalent widths of \heii\ and \hg.}
\label{res:wdheiihr}
\end{figure}

Another property to be explored is the He\,{\sc ii}/Balmer ratio which
one might expect to depend upon white dwarf mass on the basis that
accretion onto higher mass white dwarfs will produce more photons
capable of ionising He\,{\sc ii}. In Fig.~\ref{res:wdheiihr} we have
plotted those two quantities for those systems where the mass of the
white dwarf has been measured. There is no correlation apparent. In
most cases the measurements of the masses of the compact objects have
been performed by measuring the white dwarf radial velocity
semi-amplitude and combining it with the radial velocity
semi-amplitude of the companion star. Assuming a mass for the donor
star depending on its spectral type, a value for the mass of the
compact object is obtained.  It is worth mentioning that this method
is known to produce very uncertain results and that the uncertainties
in the mass determinations could completely mask any such a
relationship if it existed.

When we explore the correlation between the EW of the \heii\ line
versus the mass of the compact object, the mass of the donor star, and
the mass ratio of the system we reach the same conclusion: there is no
correlation between the masses of the components of the binaries and
the \heii\ EW. It is worth emphasising again that the values for the
white dwarf masses are very uncertain. Also we should not forget that
the EWs of the lines can change significantly for one system during
the outburst, e.g. the \heii\ EW of U~Gem during its April/May 2001
outburst changes from -1 to 4 \AA\ throughout the outburst (Steeghs
private communication).  The EW values plotted here are measured at a
particular time during outburst that is not the same for all the
systems plotted which again may mask any correlations.

\subsection{Line emission versus orbital period}

\begin{figure}
\begin{picture}(100,0)(-270,250)
\put(0,0){\includegraphics{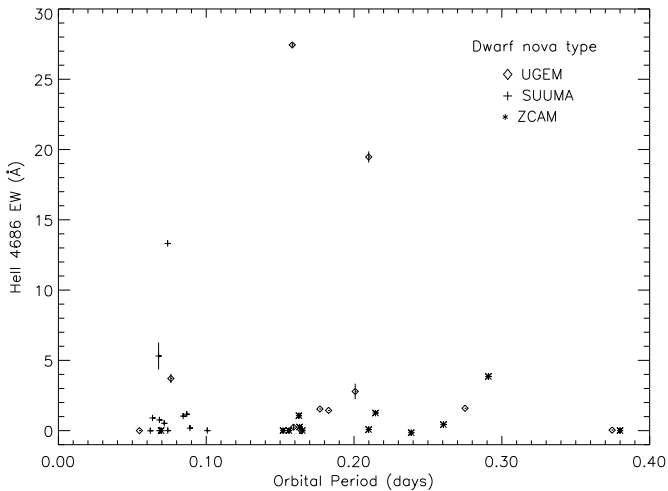}}
\noindent
\end{picture}
\vspace{75mm}
\caption{\heii\ equivalent width versus the orbital period of the
  dwarf nova. Different symbols are used for each dwarf nova type.}
\label{res:heiiewperiod}
\end{figure}

In Fig.~\ref{res:heiiewperiod} we present the EW of the \heii\ line
measured from each dwarf nova spectrum presented in this paper versus
its orbital period (Ritter \& Kolb 1998). As there is a correlation
between the dwarf nova type and its orbital period, we have used
different symbols for U~Gem, SU~UMa and Z~Cam systems. It is clear
that there is no correlation between the orbital period of the CV and
the strength of the \heii\ emission. The orbital periods of 9 of the
dwarf novae presented in this study are unknown and therefore have not
been included in this plot.

\subsection{EW and FWHM versus outburst phase}

\begin{figure}
\begin{picture}(100,0)(-270,250)
\put(0,0){\includegraphics{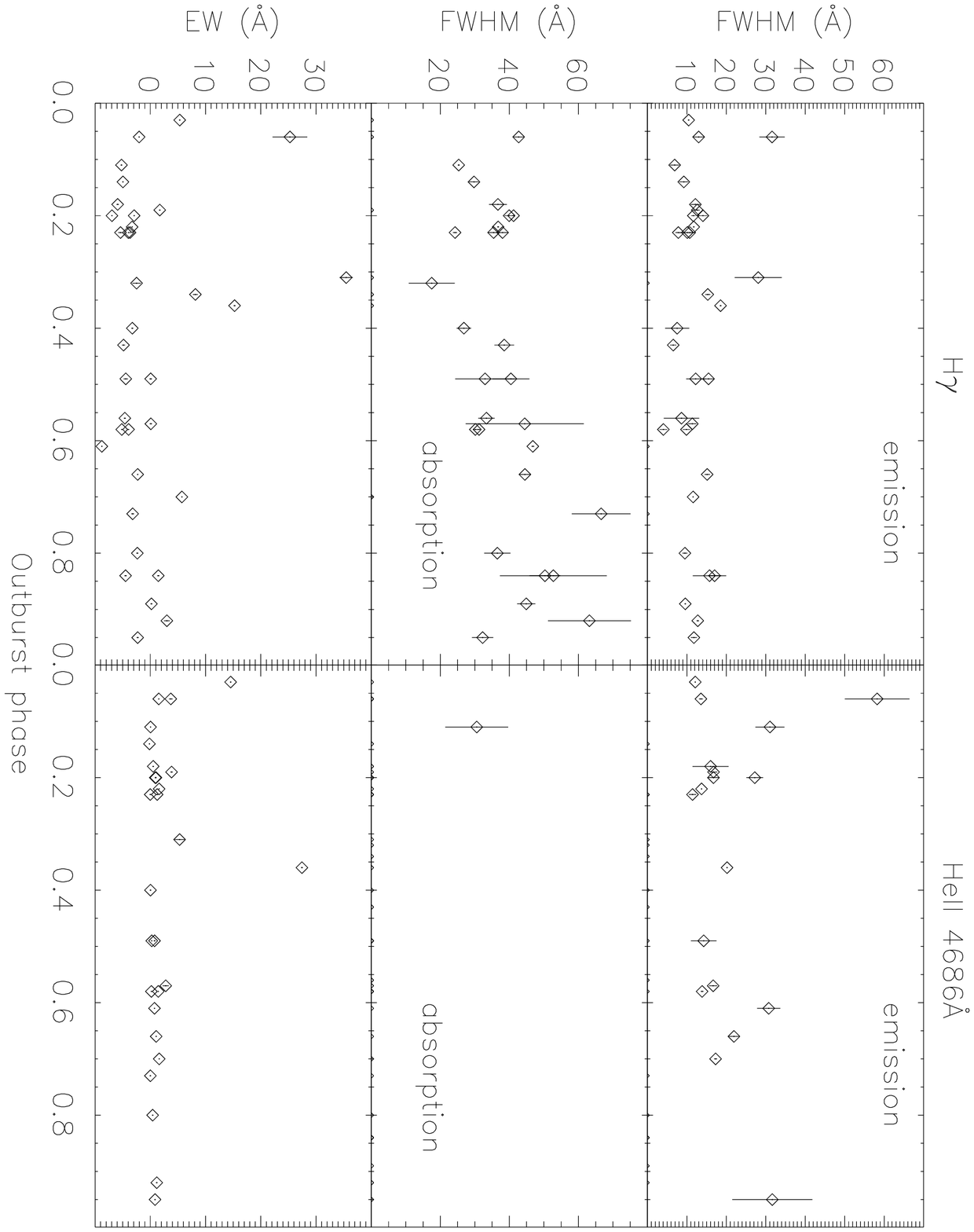}}
\put(0,0){\includegraphics{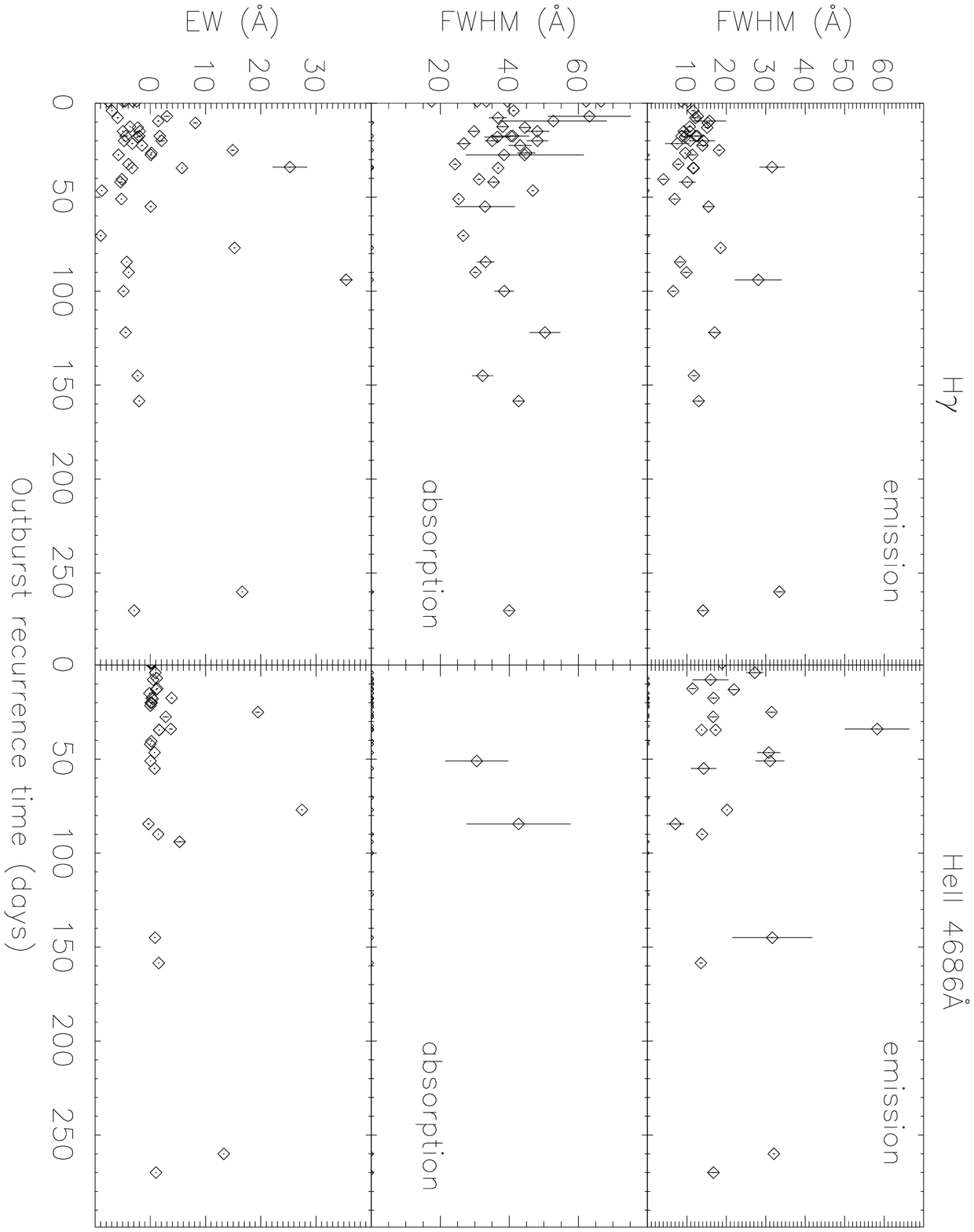}}\noindent
\noindent
\end{picture}
\vspace{140mm}
\caption{FWHM and EW for \hg\ and \heii\ versus the outburst phase at
  which the dwarf novae were observed (top panels) and versus the
  average time between outbursts (bottom panels).}
\label{res:phase}
\end{figure}

 \begin{table}
 \centering
 \caption{List of the dwarf novae and the outburst phase at
   which they were observed. The average outburst recurrence time (RT)
   in days as calculated by using VSNET data is also given.}
 \label{res:phase:tab}
 \begin{center}
 \begin{tabular}{lcclcc}
Target & RT & $\phi_{\rm out}$ & Target & RT & $\phi_{\rm out}$\\
\hline
FO~And    &8 & 0.18 &IR~Gem     &33 &0.23\\
LX~And    &41 & 0.58 &U~Gem     &159&0.06\\
RX~And    &20 &      &X~Leo     &23 &\\
FO~Aql    &22 & 0.40 &CY~Lyr    &20 &\\
VZ~Aqr    &85 &      &V419~Lyr  &350 &\\
SS~Aur    &90 & 0.58 &BI~Ori    &55 &0.49\\
AT~Cnc    &15 & 0.14 &CN~Ori    &18 &0.49\\
SY~Cnc    &27 & 0.89 &V1159~Ori &42 &0.23\\
YZ~Cnc    &7  & 0.92 &HX~Peg    &28 &0.57\\
HL~CMa    &13 & 0.23 &IP~Peg    &77 &0.36\\
SV~CMi    &28 &      &RU~Peg    &51 &0.11\\
AM~Cas    &18 &      &GK~Per    &1095&0.03\\
GX~Cas    &   &      &KT~Per    &13 &0.66\\
KU~Cas    &71 &      &TZ~Per    &18 &0.80\\
TU~Crt    &270 & 0.20 &TY~Psc   &47 &0.61\\
EM~Cyg    &18 & 0.19 &VZ~Pyx    &11 &0.34\\
SS~Cyg    &35 & 0.22,0.70 &AW~Sge  & &0.73  \\
V516~Cyg  &15 &      &RZ~Sge    &122&0.84\\
V542~Cyg  &   &      &V893~Sco  &34 &0.06\\
V792~Cyg  &145 & 0.95 &NY~Ser   & & \\
V1504~Cyg &   & 0.56 &DI~UMa    & &0.32\\
AB~Dra    & 10& 0.84 &ER~UMa    &4  &0.20\\
EX~Dra    & 25&      &IY~UMa    &260& \\ 
AH~Eri    & 100& 0.43 &SS~UMi   &94 &0.31\\
\hline
 \end{tabular}
 \end{center}
 \end{table}
 
 We already mentioned in section 4.2 that the EW of one particular
 line might change considerably throughout the outburst. By
 introducing the concept of outburst phase, $\phi_{\rm out}$, (time
 since the start of the outburst divided by the length of the
 outburst) we can use our sample of dwarf novae to study whether there
 is any relationship between the EW and FWHM of the lines and the
 outburst phase at which they were observed. The outburst phases for
 the dwarf novae observed are given in Table~\ref{res:phase:tab}. The
 top panel of Fig.~\ref{res:phase} presents the FWHM and EW measured
 for \hg\ and \heii. Two values for the FWHM of \hg\ are presented,
 one was measured when the line was in absorption and the other when
 the line was in emission or when the absorption line showed an
 emission core. The bottom panel of Fig.~\ref{res:phase} shows also
 the FWHM and EW for both lines but this time versus the average times
 between outbursts. We calculated these by using VSNET ligthcurves
 that span over years. One of the dwarf novae, GK~Per, has not been
 included in this bottom figure because its outburst recurrence time
 is $\sim$1095 days (3 years) and expanding the scale that much would
 not allow us to see the other points.
 
 It is clear from the plots that there is not any clear correlation
 between the FWHM and EW of the lines and the outburst phase or the
 average time between outbursts.

\section{Conclusions}

We present the first comprehensive sample of dwarf nova spectra during
outburst. We study the possible correlations between the lines present
in the spectra and fundamental parameters of the systems.  We find
that there is no correlation between the strength of the lines and the
masses of the components of the system or the periods of the binaries.
There seems to be some correlation between the strength of the lines
and the inclination of the system as represented by the FWHM of the
lines. We do not find any correlation between the FWHM or the EW of
the lines and either the time of the outburst at which the dwarf novae
were observed or the outburst recurrence time measured for each
system.

Of the 48 dwarf novae outburst spectra presented we find that 20
systems, 13 of them very clearly, and the other 7 possibly, show
\heii\ in emission. If models presented by Smak (2001) and Ogilvie
(2001) are correct, these are good candidates to show spiral structure
in their accretion discs during outburst. These results are even more
encouraging because four of these candidates: SS~Cyg, EX~Dra, U~Gem
and IP~Peg, have been seen to show hints of spiral structure in their
discs already (Steeghs et al. 1996; Joergens, Spruit \& Rutten 2000,
Steeghs, Harlaftis \& Horne 1997, Groot 2001). Based on recent
outburst spectra of U~Gem (Steeghs, private communication) we suggest
that in fact the \heii\ emission is either the result of irradiation
or produced by the spiral structures themselves. If its origin is
irradiation we expect the \heii\ intensity to change with disc
thickness. If the origin of \heii\ is the structures themselves, we
should see them in all dwarf novae that show \heii\ emission.  Time
resolved spectroscopy of the other candidates found in our sample will
allow us to confirm or dismiss the presence of spiral asymmetries in
their accretion discs.

\subsection*{Acknowledgements}

LM-R was supported by a PPARC post-doctoral research grant. The Isaac
Newton telescope is operated on the island of La Palma by the Isaac
Newton Group in the Spanish Observatorio del Roque de los Muchachos of
the Instituto de Astrof\'{\i}sica de Canarias. The authors would like
to thank the Variable Star Network, VSNET, and the Variable Star
Observers League in Japan, VSOLJ, for making their data available as
most of the light curves presented in this paper were produced using
their observations. In this research, we have used, and acknowledge
with thanks, data from the AAVSO International Database, based on
observations submitted to the AAVSO by variable star observers
worldwide. The reduction and analysis of the data was carried out on
the Southampton node of the STARLINK network. We also wish to thank
the referee and D. Steeghs for useful comments and L. Gonz{\' a}lez
Hern{\' a}ndez for computer support.

\begin{table*}
\vbox to220mm{\vfil Landscape table to go here
\caption{}
\vfil}
\end{table*}


\begin{thebibliography}{}

\bibitem[\protect\citename{Baba, Sadakane \& Norimoto}2001]{bsn01}
  Baba\,H., Sadakane\,K., Norimoto\,Y., 2001, IAUC 7672

\bibitem[\protect\citename{Balbus \& Hawley}1991]{bh91} Balbus\, S.,
  Hawley\, J., 1991, ApJ, 376, 214

\bibitem[\protect\citename{Downes, Webbink \& Shara}1997]{dws97}
  Downes\,R.\,A., Webbink\,R.\,D., Shara\,M.\,M., 1997, PASP, 109, 345

\bibitem[\protect\citename{Groot}2001]{g01}Groot\,P., 2001, ApJL, 551, L89
  
\bibitem[\protect\citename{Harlaftis et al. }1999]{h99b}Harlaftis\,
  E.\,T., Steeghs\, D., Horne\, K., Mart\'{\i}n\, E., Magazz\'{u}\,
  A., 1999, MNRAS, 306, 348

\bibitem[\protect\citename{Horne }1986]{h86} Horne\, K., 1986,
  PASP, 98, 609

\bibitem[\protect\citename{Joergens, Spruit \& Rutten}2000]{jsr00}
  Joergens\,V., Spruit\,H.\,C., Rutten\,R.\,G.\,M., 2000, A\&A, 356, L33
  
\bibitem[\protect\citename{Kuulkers et al.}2000]{k02} Kuulkers\,E.,
  Knigge\,C., Steeghs\,D., Wheatley\,P., Long\,K.\,S., 2002, to appear
  in the ASP Conference Series, The Physics of Cataclysmic Variables
  and Related Objects, Eds B. T. G{\"a}nsicke, K. Beuermann, K.
  Reinsch (astro-ph/0110064)

\bibitem[\protect\citename{Matsumoto, Mennickent \& Kato}2000]{mmk00}
  Matsumoto\, K., Mennickent\,R.\,E., Kato\,T., 2000, A\&A, 363, 1029

\bibitem[\protect\citename{Morales-Rueda, Marsh \&
    Billington}2000]{mmb00} Morales-Rueda L., Marsh T.\,R.,
    Billington I., 2000, MNRAS, 313, 454
    
\bibitem[\protect\citename{Morales-Rueda et al.}1999]{m99}
    Morales-Rueda\,L., Still\,M.\,D., Roche\,P., 1999, MNRAS, 306, 753

\bibitem[\protect\citename{Morales-Rueda et al.}2001]{m01}
    Morales-Rueda\,L., Still\,M.\,D., Roche\,P., Wood\,J.\,H.,
    Lockley\,J.\,J., 2002, MNRAS, 329, 597
  
\bibitem[\protect\citename{Nogami et al.}1999]{n99} Nogami\,D.,
  Masuda\,S., Kato\,T., Hirata\,R., 1999, PASJ, 51, 115

\bibitem[\protect\citename{North et al.}2000]{n00} North\,R.\,C.,
  Marsh\,T.\,R., Moran\,C.\,K., Kolb\,U., Smith\,R.\,C., Stehle\,R.,
  2000, MNRAS, 313, 383

\bibitem[\protect\citename{Ogilvie}2002]{o02} Ogilvie\,G.\,I., 2002,
  MNRAS, in press (astro-ph/0111262)

\bibitem[\protect\citename{Oke}1990]{o90} Oke\,J.\,B., 1990, AJ, 99,
  1621

\bibitem[\protect\citename{Oke \& Gunn}1983]{og83} Oke\,J.\,B.,
  Gunn\,J.\,E., 1983, ApJ, 266, 713

\bibitem[\protect\citename{Osaki }1974]{o74} Osaki\, Y., 1974, PASJ, 26, 429

\bibitem[\protect\citename{Patterson et al.}2000]{p00} Patterson\,J.,
  Kemp\,J., Jensen\,L., Vanmunster\,T., Skillman\,D.\,R., Martin\,B.,
  Fried\,R., Thorstensen\,J.\,R., 2000, PASP, 112, 1567

\bibitem[\protect\citename{Ritter \& Kolb}1998]{rk98} Ritter\,H.,
Kolb\, U., 1998, A\&AS, 129, 83

\bibitem[\protect\citename{Sawada, Matsuda \& Hachisu }1986]{smh86}
  Sawada\, E., Matsuda\, T., Hachisu\, I., 1986, MNRAS, 219, 75
  
\bibitem[\protect\citename{Steeghs, Harlaftis \& Horne }1997]{shh97}
  Steeghs\, D., Harlaftis\, E.\, T., Horne\, K., 1997, MNRAS, 270, L28

\bibitem[\protect\citename{Steeghs et al. }1996]{s96} Steeghs\, D.,
  Horne\, K., Marsh\, T.\, R., Donati\, J.\, F., 1996, MNRAS, 281,
  626

\bibitem[\protect\citename{Szkody et al.}1999]{s99} Szkody P., Linnell
  A., Honeycutt K., Robertson J., Silber A., Hoard D.\,W., Pastwick
  L., Desai V., Hubeny I., Cannizzo J., Liller W., Zissell R., Walker
  G., 1999, ApJ, 521, 362


\bibitem[\protect\citename{Smak}2001]{s01} Smak\,J.\,I., 2001, Acta
  Astron., 51, 295

\bibitem[\protect\citename{Thorstensen}1997]{t97} Thorstensen J.\,R.,
  1997, PASP, 109, 1241

\bibitem[\protect\citename{Uemura et al}2000]{u00} Uemura M., et al.,
  2000, PASJ, 52, L9

\bibitem[\protect\citename{Warner}1995]{w95} Warner\,B.,
  1995, Cambridge Astrophysics Series 28. Cataclysmic Variable
  Stars. Cambridge Univ. Press, Cambridge

\end{thebibliography}
\end{document}